\documentclass[12pt,a4paper]{article}
\usepackage{amsmath}
\usepackage{amssymb}
\usepackage{cite}

\renewcommand{\thesection}
  {\arabic{section}.\hspace{-.5em}}

\makeatletter
\renewcommand\section{
  \@startsection{section}{3}{\z@}%
  {-3.25ex\@plus -1ex \@minus -.2ex}%
  {1.5ex \@plus .2ex}%
  {\normalfont\normalsize\bfseries\mathversion{bold}}}
\renewcommand\subsection{
  \@startsection{subsection}{3}{\z@}%
  {-3.25ex\@plus -1ex \@minus -.2ex}%
  {1.5ex \@plus .2ex}%
  {\normalfont\normalsize\bfseries\mathversion{bold}}}
\makeatother

\makeatletter \@addtoreset{equation}{section} \makeatother
\renewcommand{\theequation}{\arabic{section}.\arabic{equation}}

\makeatletter
\renewcommand{\appendix}{
\renewcommand{\thesection}{\Alph{section}.\hspace{-.5em}}
\@addtoreset{equation}{section}
\renewcommand{\theequation}{\Alph{section}.\arabic{equation}}
\setcounter{section}{0}}
\makeatother

\let\oldthebibliography\thebibliography
\renewcommand\thebibliography[1]{
  \oldthebibliography{#1}\setlength{\itemsep}{0.4ex}}

\newcommand{\atmark}{{\scriptsize
  \textcircled{\raisebox{-.15ex}{\small\it\hspace{-.12em}a}}}}

\newcommand{\Eqn}[1]{&\hspace{-0.5em}#1\hspace{-0.5em}&}
\newcommand{\nn}{\nonumber}
\renewcommand{\[}{\begin{eqnarray}}
\renewcommand{\]}{\end{eqnarray}}

\newcommand{\alg}[1]{\mathfrak{#1}}
\newcommand{\grp}[1]{\mathrm{#1}}
\newcommand{\bvec}[1]{\boldsymbol{#1}}
\newcommand{\varth}{\vartheta}

\newcommand{\bbR}{{\mathbb R}}
\newcommand{\bbZ}{{\mathbb Z}}

\newcommand{\scr}[1]{{\mbox{\scriptsize #1}}}
\newcommand{\sscr}[1]{{\mbox{\tiny #1}}}

\newcommand{\arctanh}{\mathop{\rm arctanh}}

\begin{document}


\def\papertitlepage{\baselineskip 3.5ex \thispagestyle{empty}}
\def\preprinumber#1#2{\hfill
\begin{minipage}{1.2in}
#1 \par\noindent #2
\end{minipage}}

%
\papertitlepage
\setcounter{page}{0}
\preprinumber{arXiv:1603.09108}{}
\vskip 2ex
\vfill
\begin{center}
{\large\bf\mathversion{bold}
BPS index and 4d ${\cal N}=2$ superconformal field theories
}
\end{center}
\vfill
\baselineskip=3.5ex
\begin{center}
Kazuhiro Sakai\\

{\small
\vskip 6ex
{\it Institute of Physics, Meiji Gakuin University,
Yokohama 244-8539, Japan}\\
\vskip 1ex
{\tt kzhrsakai\atmark gmail.com}

}
\end{center}
\vfill
\baselineskip=3.5ex
\begin{center} {\bf Abstract} \end{center}

We study the BPS index for the four-dimensional rank-one
${\cal N}=2$ superconformal field theories $H_0,H_1,H_2,E_6,E_7,E_8$.
We consider compactifications of the E-string theory on $T^2$
in which these theories arise as low energy limits.
Using this realization we clarify the general structure
of the BPS index. The index is characterized by two exponents
and a sequence of invariants. We determine the exponents
and the first few invariants.

\vfill
\noindent
March 2016


\setcounter{page}{0}
\newpage
\renewcommand{\thefootnote}{\arabic{footnote}}
\setcounter{footnote}{0}
\setcounter{section}{0}
\baselineskip = 3.5ex
\pagestyle{plain}
%

\section{Introduction and summary}

The moduli space of vacua of ${\cal N}=2$ supersymmetric theories
in four dimensions often contains singularities
where a nontrivial superconformal field theory (SCFT) arises
\cite{Argyres:1995jj}.
Nowadays many such 4d ${\cal N}=2$ SCFTs are known,
even in the rank-one case alone
(see, e.g., \cite{Argyres:2016xua}).
A classic example is the sequence of SCFTs
denoted by $H_0,H_1,H_2,D_4,E_6,E_7,E_8$
\cite{Argyres:1995xn,Seiberg:1994aj,Minahan:1996fg,Minahan:1996cj},
which can be realized by a single D3-brane probing
F-theory singularities with constant dilaton
\cite{Vafa:1996xn,Morrison:1996na,Morrison:1996pp,
Sen:1996vd,Banks:1996nj,Dasgupta:1996ij}.

In the study of 4d ${\cal N}=2$ supersymmetric theories
the BPS index is of crucial importance
as it captures the details of the exact quantum spectrum.
For $\grp{U}(n)$ gauge theories
the BPS index can be expressed in a very concise, explicit form
known as the Nekrasov partition function
\cite{Nekrasov:2002qd,Nekrasov:2003rj}.\footnote{
In this paper the term `Nekrasov partition function' means
the explicit expression (typically given as a sum over partitions),
while the term `BPS index' abstractly denotes
the observable that can, in certain cases, be expressed as
the Nekrasov partition function.
For 5d gauge theories on $\bbR^4\times S^1$
the BPS index is defined as a generalized supersymmetric index
\cite{Nekrasov:2002qd}.
For 4d gauge theories it can be defined as
the partition function
given by the path integral in the Omega background
\cite{Nekrasov:2003rj}.
For the SCFTs studied in this paper we do not know
any explicit, direct definition of the BPS index.
It can, however, be defined indirectly
by means of theories which flow to the SCFTs
and one can unambiguously compute it
(at least as a series expansion).}
It is natural to ask how the BPS index looks near
the SCFT singularities on the moduli space of supersymmetric theory.
The answer is not obvious, even in the case of a gauge theory
whose Nekrasov partition function is explicitly known.
This is because the Nekrasov partition function is given by the
instanton expansion about the classical singularity of the moduli space,
but the expansion is no longer valid at the SCFT singularities.
Some sort of analytic continuation is required.

In this paper we focus on the 4d rank-one ${\cal N}=2$ SCFTs
$H_n,E_{8-n}\ (n=0,1,2)$ and study the BPS index.
(The index for the $D_4$ SCFT can be computed from the Nekrasov
partition function by using the modular properties of the index
\cite{Grimm:2007tm}.)
These SCFTs are realized in many ways.
For example, the $H_n$ SCFTs can be studied by means of
the original realization 
in $\grp{SU}(2)$ super Yang--Mills \cite{Argyres:1995xn}.
Analysis of the BPS index along these lines was carried out
\cite{Huang:2009md}.
In this paper we study the SCFTs by means of 
toroidal compactifications of the 6d E-string theory
\cite{Ganor:1996mu,Seiberg:1996vs,Klemm:1996hh,
Ganor:1996pc,Minahan:1998vr,Eguchi:2002fc,Kim:2014dza}.
This approach has the following advantages.

Firstly, one can make full use of the known results
of the well-studied BPS index of E-strings.
It has several entirely different interpretations,
providing us with complementary ways to compute it.
In particular, the E-string theory has a world-sheet description,
which enables us to compute the index as the generating function
for the sequence of elliptic genera of
multiple E-strings \cite{Kim:2014dza}.
Secondly, in this approach one can study the above SCFTs
(including the $D_4$ case \cite{Sakai:2014hsa})
in a unified manner. The E-string theory
encompasses almost all 5d ${\cal N}=1$ and 4d ${\cal N}=2$
rank-one gauge theories \cite{Ganor:1996pc, Eguchi:2002fc}.
All the above SCFTs arise in the moduli space of
this single theory.

We clarify the general structure of the BPS index.
For the sake of simplicity we consider the unrefined case,
where the chemical potentials $\epsilon_1,\epsilon_2$
for the Lorentz spins are fixed as
$\epsilon_1=-\epsilon_2=:\hbar$.
Our main results are summarized as follows.
The BPS index for the type $\alg{g}=E_{8-n},H_n\ (n=0,1,2)$ SCFTs
takes the form
\[
Z^\alg{g}(\phi,\hbar)
=\exp\left(\beta^\alg{g}\hbar^2\partial_\phi^2\right)
\left[
  \phi^{-\gamma^\alg{g}}
  \exp\sum_{k=1}^\infty
  c^\alg{g}_k\left(\frac{\hbar}{\phi}\right)^{m^\alg{g} k}
\right],
\]
where $\phi$ is the Higgs expectation value and
\[
\beta^{E_{8-n}}\Eqn{=}\beta^{H_n}=
  \frac{3^{(n-1)/2}}{4\pi},\\
\gamma^{E_{8-n}}\Eqn{=}\frac{1}{2}\left(\frac{12}{n+2}-1\right),
\qquad
\gamma^{H_n}=\frac{1}{2}\left(\frac{12}{n+2}-1\right)^{-1},\\[1ex]
m^{E_{8-n}}\Eqn{=}m^{H_n}=4+2|n-1|.
\]
The differential operator is introduced so that
the `descendants' that are determined by the modular anomaly equation
are concealed from view.
Put in this form the BPS index is characterized by
two exponents $\beta^\alg{g}, \gamma^\alg{g}$
and a sequence of invariants $c^\alg{g}_k$.
In this paper we determine
$\beta^\alg{g}, \gamma^\alg{g}$ (as above), invariants
$c^{E_8}_k,c^{H_0}_k\ (k=1,2,3,4)$ and $c^{E_7}_1,c^{H_1}_1$
by using the known results of the BPS index for the E-string theory.
The values of $\gamma^{H_n}$ have been known \cite{Huang:2009md}
and our results are in agreement with them.

Our method can easily be generalized to the
case with general $\epsilon_1,\epsilon_2$.
Another interesting generalization of this work is to turn on
the chemical potentials for the other global symmetry charges.
It would also be interesting to clarify
how our results are related to the superconformal index
for the SCFTs
\cite{Gadde:2010te,Gadde:2011uv,Buican:2015ina,Buican:2015hsa,
Cordova:2015nma,Cecotti:2015lab}.

The rest of the paper is organized as follows.
In section~2 we review the definition and the basic properties
of the BPS index of E-strings.
In section~3 we consider the setting of the E-string theory
on $\bbR^4\times T^2$ that realizes $E_8\oplus H_0$ singularities
on the moduli space.
We first study the index of E-strings in this setting and then
take limits to obtain the index for the $E_8$ and $H_0$ SCFTs.
Section~4 and section~5 are devoted to the
$E_7\oplus H_1$ case and the $E_6\oplus H_2$ case respectively.

\section{Review of BPS index of E-strings}

The BPS index of E-strings is defined as
the 5d BPS index \cite{Nekrasov:2002qd}
for the E-string theory on $\bbR^5\times S^1$.
It is given by a trace over the space of the BPS particles as follows:
\[\label{Zgendef}
Z(\phi,\tau,\bvec{m},\epsilon_1,\epsilon_2)
:= {\rm Tr}\, (-1)^{2J_\sscr{L}+2J_\sscr{R}}
 y_\scr{L}^{J_\sscr{L}}
 y_\scr{R}^{J_\sscr{R}+J_\sscr{I}}
 p^n q^k e^{i\bvec{\Lambda}\cdot\bvec{m}},
\]
where
\[
y_\scr{L} := e^{i(\epsilon_1-\epsilon_2)},\qquad
y_\scr{R} := e^{i(\epsilon_1+\epsilon_2)},\qquad
p := e^{-\phi},\qquad
q := e^{2\pi i\tau}.
\]
Here $J_\scr{L},J_\scr{R},J_\scr{I}$ and
$\bvec{\Lambda}=(\Lambda_1,\dots,\Lambda_8)$ are spins
(or weights of the associated Lie algebras) of the little group
$\grp{SO}(4)=\grp{SU}(2)_\scr{L}\times\grp{SU}(2)_\scr{R}$,
the R-symmetry group $\grp{SU}(2)_\scr{I}$ and the global
symmetry group $E_8$ respectively. Nonnegative integers $n,k$
are respectively the winding number and the momentum along $S^1$.
$Z$ is a function in twelve variables.
$\phi$ is the tension of the E-strings and
$\tau$ is proportional to
the inverse of the radius of $S^1$.
In the Seiberg--Witten description
\cite{Seiberg:1994rs,Seiberg:1994aj}
of the E-string theory on $\bbR^4\times T^2$,
$\phi$ is interpreted as the Higgs expectation value of
the $\grp{U}(1)$ vector multiplet
and $\tau$ is the complex structure of $T^2$.
$\bvec{m}=(m_1,\ldots,m_8)$ and $\epsilon_1,\epsilon_2$ are
respectively the Wilson line parameters or the chemical potentials
for the global symmetries $E_8$ and $\grp{SO}(4)$.
Throughout this paper we consider the unrefined case
$\epsilon_1=-\epsilon_2=:\hbar$.

The index $Z$ is interpreted in several ways.
One interpretation is associated with the expansion
\[\label{pexp}
Z\Eqn{=}1+\sum_{n=1}^\infty p^nZ_n,
\]
where $Z_n$ is the elliptic genus of $n$ E-strings.
$Z_n$ with any $n$ can in principle be computed
by using the localization technique \cite{Kim:2014dza}.
Explicit forms of $Z_n\ (n\le 4)$ were obtained in \cite{Kim:2014dza}.
Another useful interpretation is
\[
Z\Eqn{=}\exp\sum_{g=0}^\infty \hbar^{2g-2}F_g,
\]
where $F_g$ is the genus-$g$ topological string amplitude
for the local $\frac{1}{2}$K3.
$F_g\ (g\le 3)$ with general $\bvec{m}$ were computed
in \cite{Sakai:2011xg},
which we will use mainly in this paper.
A third interpretation relates $Z$ with
the partition function of a certain
five-brane web system \cite{Kim:2015jba}.
This picture
enables us to compute $Z$ as a power series expansion in $q$.

The index $Z$ satisfies two important constraints.
One is known as the modular anomaly equation
\cite{Hosono:1999qc}
\[\label{MAE}
\partial_{E_2}Z
=\frac{1}{24}\hbar^2\partial_\phi\left(\partial_\phi-1\right)Z,
\]
where $\partial_{E_2}$ is the formal derivative in
Eisenstein series $E_2(\tau)$.
The other is the gap condition
\[\label{gapcond}
\ln Z=\sum_{n=1}^\infty p^n
  \left(\frac{1}{n\left(2\sin\frac{n\hbar}{2}\right)^2}
    +{\cal O}\left(q^n\right)\right).
\]
This follows from the geometric structure of the
local $\frac{1}{2}$K3 \cite{Minahan:1998vr}.

\section{$E_8\oplus H_0$ case}

\subsection{Seiberg--Witten curve}

Let us first consider the E-string theory on $\bbR^4\times T^2$
without $E_8$ Wilson lines,
i.e.
\[
\bvec{m}=\bvec{0}.
\]
The low energy effective theory is characterized
by the following elliptic
Seiberg--Witten curve \cite{Ganor:1996pc, Eguchi:2002fc}
\[\label{mE8curve}
y^2=4x^3-\frac{1}{12}E_4(\tau)u^4x-\frac{1}{216}E_6(\tau)u^6+4u^5.
\]
Here $u$ parametrizes the Coulomb branch of
the moduli space of vacua.
In this section we focus on the case where
the complex structure of the torus is fixed to the special value
\[
\tau = \tau_* :=  e^{2\pi i/3}.
\]
This is a value for which Eisenstein series $E_4(\tau)$ vanishes
\[
E_{4*}:=E_4(\tau_*)=0.
\]
We let subscript $*$ denote that
the quantity is evaluated at $\tau=\tau_*$.
For example,
\[
q_*
 = e^{2\pi i\tau_*}
 =-e^{-\pi\sqrt{3}}.
\]
Other Eisenstein series take the following values
(see, e.g., \cite{Abramowitz})
\[
E_{2*}=\frac{2\sqrt{3}}{\pi},\qquad
E_{6*}=216\frac{\Gamma(1/3)^{18}}{(2\pi)^{12}}.
\]
The Seiberg--Witten curve (\ref{mE8curve}) becomes
\[
y^2\Eqn{=}4x^3-\frac{1}{216}E_{6*}u^6+4u^5\\
   \Eqn{=}4x^3-\frac{\Gamma(1/3)^{18}}{(2\pi)^{12}}u^6+4u^5.
\]
Because of the absence of the linear term in $x$,
the $j$-invariant of this elliptic curve is identically $0$.
Therefore the complex structure of the curve takes
the constant value $\tau_*$ over the moduli space.
The discriminant of the curve has zeros
at $u=0$ and $u=864/E_{6*}$,
where the elliptic fibration has singular fibers
of Kodaira type II$^*$ and II respectively.
This means that
near the point at $u=0$ the theory looks like
the $E_8$ SCFT while near the point at $u=864/E_{6*}$
the theory looks like the $H_0$ SCFT \cite{Ganor:1996pc}.
In the following, we will study how the BPS index looks
near these SCFT points.

\subsection{Mirror map}

In this subsection we establish the mirror map
which connects the global moduli space coordinate $u$
with the scalar expectation value $\phi$ on local patches
of the moduli space. We will see that
the mirror map reduces to a very simple form
in the present case with $\tau=\tau_*$.
This is similar to what was observed in
the $D_4\oplus D_4$ case \cite{Sakai:2014hsa}.

Since the mirror map described below is merely
a special case of the general one \cite{Eguchi:2002fc},
we will summarize the main points only.
The reader is referred for the details
to \cite{Sakai:2011xg}. Note that the variable
$\phi$ used in \cite{Sakai:2011xg} should not be confused
with $\phi$ in this paper. They are related to each other by
\[
\phi_\scr{there}
 =-\phi_\scr{here}
  +\ln\left(-q\prod_{k=1}^\infty(1-q^k)^{12}\right).
\]

In our present case with $\bvec{m}=\bvec{0}$ and $\tau=\tau_*$,
one of the periods (divided by $2\pi$) of the Seiberg--Witten curve
is written as
\[
\omega=\frac{1}{u\left(1-\frac{864}{E_{6*} u}\right)^{1/6}}
 =\frac{E_{6*}}{864}\frac{z^6-1}{z^5}.
\]
Here $z$ is a new coordinate of the moduli space defined by
\[
z:=\left(1-\frac{864}{E_{6*}u}\right)^{-1/6}.
\]
The singularities corresponding to the $E_8$ and $H_0$ SCFTs
are mapped to $z=0$ and $z=\infty$ respectively.
The Higgs expectation value is given by
\[
\phi\Eqn{=}\int\omega du\nn\\
 \Eqn{=}6\int\frac{dz}{1-z^6}\\
 \Eqn{=}\ln\frac{1+z}{1-z}+\frac{1}{2}\ln\frac{1+z+z^2}{1-z+z^2}
  +\sqrt{3}\arctan\left(\frac{\sqrt{3}z}{1-z^2}\right)+\phi_0^{E_8}.
\]
Here $\phi^{E_8}_0$ is an integration constant.
Near the $E_8$ singularity at $z=0$,
$\phi$ is expanded as
\[\label{phiexpE8}
\phi-\phi_0^{E_8}
 =6\sum_{k=0}^\infty \frac{z^{6k+1}}{6k+1}
 =6\left(z+\frac{z^7}{7}+\frac{z^{13}}{13}+\cdots\right).
\]
Near the $H_0$ singularity at $z=\infty$, on the other hand,
$\phi$ is expanded as
\[\label{phiexpH0}
\phi-\phi_0^{H_0}
 =6\sum_{k=0}^\infty\frac{1}{(6k+5)z^{6k+5}}
 =6\left(\frac{1}{5z^5}+\frac{1}{11z^{11}}
        +\frac{1}{17z^{17}}+\cdots\right).
\]
It is possible to determine the values of
the integration constants $\phi_0^{E_8}$ and $\phi_0^{H_0}$
with respect to the convention of \cite{Sakai:2011xg},
but for our purposes their values are not important.

\subsection{6d amplitudes}

Let us now consider the BPS index of E-strings
in the present setting
\[
Z^{E_8\oplus H_0}(\phi,\hbar)
:=Z
(\phi,\tau=\tau_*,\bvec{m}=\bvec{0},\epsilon_1=\hbar,\epsilon_2=-\hbar).
\]
We will study it mainly in the expansion of the form
\[
\ln Z^{E_8\oplus H_0}
 =\sum_{g=0}^\infty \hbar^{2g-2} F^{E_8\oplus H_0}_g(\phi).
\]
By using the modular anomaly equation (\ref{MAE}),
the gap condition (\ref{gapcond}),
the known forms of $F_g\ (g\le 3)$ \cite{Sakai:2011xg}
and elliptic genera $Z_n\ (n\le 4)$ \cite{Kim:2014dza},
we are able to determine $F^{E_8\oplus H_0}_g$ for $g\le 15$.
The first few amplitudes are as follows:
\[
F^{E_8\oplus H_0}_0\Eqn{=}0,\\
F^{E_8\oplus H_0}_1\Eqn{=}\frac{1}{2}\ln\omega+\frac{\phi}{2}
  -\frac{1}{2}
   \ln\left(-q_*\prod_{k=1}^\infty(1-q_*^k)^{12}\right)\nn\\
 \Eqn{=}
  \frac{1}{2}\ln\frac{z^6-1}{z^5}+\frac{\phi}{2}
  +\frac{\sqrt{3}}{4}\pi-\ln 2-\frac{3}{4}\ln 3,\\
F^{E_8\oplus H_0}_2
  \Eqn{=}
  \left(\frac{35-10z^6+11z^{12}}{3456z^{2}}
       -\frac{1}{96}\right)E_{2*},\\
F^{E_8\oplus H_0}_3
  \Eqn{=}\frac{5(1-z^6)^2(14+23z^6+44z^{12})}{746496 z^4}E_{2*}^2,\\[1ex]
F^{E_8\oplus H_0}_4\Eqn{=}\frac{(1-z^6)^2}{z^6}
 \left[
 \frac{5(2485+3128z^6+3246z^{12}-31000z^{18}+92125z^{24})}
      {7739670528}E_{2*}^3\right.\nn\\
&&\hspace{3em}\left.
 +\frac{12625-34792z^6+632886z^{12}-2352376z^{18}+2208217z^{24}}
       {38698352640}E_{6*}
 \right].\nn\\
\]
The rest of the results are rather lengthy
and thus we do not present them here.
Instead, in what follows we clarify
the structure of $F^{E_8\oplus H_0}_g$ with general $g$.

First, $F^{E_8\oplus H_0}_g\ (g\ge 3)$ takes the
following form
\[\label{FE8H0genform}
F^{E_8\oplus H_0}_g=\frac{(1-z^6)^2}{z^{2g-2}}
 \sum_{k=0}^{\lfloor (g-1)/3\rfloor}
E_{2*}^{g-1-3k}E_{6*}^{k}\sum_{i=0}^{2g-4}c_{g,k,i}z^{6i},
\]
where $c_{g,k,i}$ are some numerical coefficients.
Next, one can easily see that $F^{E_8\oplus H_0}_g$ with
$g=3n+2,3n+3\ (n\in\bbZ_{\ge 0})$ are completely determined
by the modular anomaly equation, given the data of
$F^{E_8\oplus H_0}_g\ (g\le 3n+1)$.
Therefore all the essential data are provided by
$F^{E_8\oplus H_0}_g$ with $g=3n+1\ (n\in\bbZ_{\ge 0})$.
Furthermore, apart from the $n=0$ case, they are written as
\[
F^{E_8\oplus H_0}_{3n+1}
=\frac{(1-z^6)^2}{z^{6n}}
 \sum_{k=0}^n E_{2*}^{3(n-k)}E_{6*}^{k}
 \sum_{i=0}^{6n-2}c_{3n+1,k,i}z^{6i}
\]
with unknowns $c_{3n+1,k,i}$,
but except for $k=n$ they are again determined by
the modular anomaly equation,
given the data of $F^{E_8\oplus H_0}_{3m+1}\ (m<n)$.
To sum up, the data of $F^{E_8\oplus H_0}_1$
and $c_{3n+1,n,i}\ (n\in\bbZ_{>0},\ i=0,\ldots,6n-2)$
completely determine the BPS index.

Before closing this subsection let us sketch out a proof
of the general form (\ref{FE8H0genform}).
First, it is well-known that
topological string amplitudes for Calabi--Yau threefolds
are polynomials in a finite number of generators
\cite{Yamaguchi:2004bt,Alim:2007qj}.
The generators for the most general BPS index of E-strings
were identified \cite{Sakai:2011xg}.
By reducing the results to the present case,
one sees that
$F^{E_8\oplus H_0}_g\ (g\ge 2)$ are polynomials in
$\partial_\phi^k\ln\omega \ (k=1,2,3)$ and in $E_{2*},E_{6*}$.
Moreover, $F^{E_8\oplus H_0}_g\ (g\ge 3)$ is of degree $2g-2$
and weight $2g-2$, where we assign degree $k$ to
$\partial_\phi^k\ln\omega$ and weight $2n$ to $E_{2n*}$.
It then follows that $F^{E_8\oplus H_0}_g\ (g\ge 3)$,
as a function in $z$, is written as
\[\label{FgE8zpoly}
F^{E_8\oplus H_0}_g=\frac{f_{2g-2}(z^6)}{z^{2g-2}},
\]
with $f_{2g-2}(x)$ being a degree $2g-2$ polynomial function.
Next, let us show that $F^{E_8\oplus H_0}_g\ (g\ge 3)$
contains the factor $(1-z^6)^2$.
We first note that the elliptic genus
of a single E-string in the present case vanishes
\[
Z^{E_8\oplus H_0}_1
 =q_*^{1/2}\frac{E_{4*}}{\eta(\tau_*)^6\varth_1(\hbar,\tau_*)^2}=0.
\]
In terms of $F^{E_8\oplus H_0}_g$ this means that
\[\label{FgE8H0gap}
F^{E_8\oplus H_0}_g={\cal O}(p^2)
\]
in the power series expansion in $p=e^{-\phi}$.
Since the variable $z$ is expanded in $p$ as
\[
z=1+2\sqrt{3}e^{-\pi\sqrt{3}/2}p+30e^{-\pi\sqrt{3}}p^2
 +{\cal O}\left(p^3\right),
\]
$F^{E_8\oplus H_0}_g\ (g\ge 2)$ has to contain the factor $(1-z)^2$
in order to satisfy (\ref{FgE8H0gap}).
This, combined with (\ref{FgE8zpoly}), means that
$F^{E_8\oplus H_0}_g\ (g\ge 3)$ has to contain the factor $(1-z^6)^2$.

\subsection{4d limits}

We have so far studied 6d theory on $\bbR^4\times T^2$.
Let us now consider scaling limits in which
the torus $T^2$ shrinks to zero size.
To do this we first recover the length scale $R$,
which is proportional to the radii of the $T^2$.
We then take the limit of $R\to 0$ while keeping
the complex structure of the $T^2$ to be $\tau=\tau_*$.
In the low energy Seiberg--Witten description,
this procedure corresponds to zooming in on a point in the moduli space.
In particular, by suitably zooming in on the singularity
at $z=0$ ($z=\infty$), one obtains the $E_8$ ($H_0$) SCFT.

Let us first consider the scaling limit in which
only the local structure of the $E_8$ singularity at $z=0$
contributes to the physics.
This is achieved by first replacing
the variables as
\[
\phi-\phi_0^{E_8}\to R\phi,\qquad
z\to Rz,\qquad
\hbar\to R\hbar
\]
and then taking the limit of $R\to 0$.
In this limit a function in $z$ is dominated by
the leading order term of its expansion in $z$.
The mirror map (\ref{phiexpE8}) is simplified as
\[
\phi=6z.
\]
The BPS index for the 4d $E_8$ SCFT is obtained
from that for the E-string theory as
\[\label{ZE8def}
Z^{E_8}(\phi,\hbar)
=
({\rm const.})
\lim_{R\to 0}Z^{E_8\oplus H_0}\left(R\phi+\phi_0^{E_8},R\hbar\right)
 \Bigm|_{z\sim 0}.
\]
The normalization constant is fixed as follows.
$Z^{E_8}$ is expanded in $\hbar$ as
\[
\ln Z^{E_8}\Eqn{=}\sum_{g=0}^\infty F^{E_8}_g\hbar^{2g-2}
\]
with
\[
F^{E_8}_0\Eqn{=}0,\qquad
F^{E_8}_1=-\frac{5}{2}\ln\phi,\qquad
F^{E_8}_2=\frac{35E_{2*}}{96}\frac{1}{\phi^2},\qquad
F^{E_8}_3=\frac{35E_{2*}^2}{288}\frac{1}{\phi^4},\nn\\
F^{E_8}_4\Eqn{=}\frac{25(497E_{2*}^3+101E_{6*})}{165888}
  \frac{1}{\phi^6},\qquad\cdots.
\]
We fix the normalization of $Z^{E_8}$ so that
$F^{E_8}_1$ takes the above simple form without constant term.
Note also that the form of $F^{E_8}_0$ is not characteristic of
the SCFT itself, but rather of how it is embedded in the bulk theory.
The explicit forms of $F^{E_8}_g\ (g\le 15)$
are immediately obtained from the results of the last subsection.
Instead of listing them all,
we will present the results in a very concise form,
taking account of the fact that 
the BPS index satisfies the modular anomaly equation.

For $Z$ that has the structure (\ref{pexp}),
one can show that the modular anomaly equation (\ref{MAE})
is formally solved as
\[\label{MAEsol}
Z=\exp\left(\frac{\phi}{2}-\frac{E_2}{96}\hbar^2\right)
  \exp\left(\frac{E_2}{24}\hbar^2\partial_\phi^2\right)\tilde Z,
\]
where $\tilde Z$ is independent of $E_2$.
After we set $\tau=\tau_*$,
$E_2$ becomes a numerical constant and
the modular anomaly equation does not make sense,
but the structure (\ref{MAEsol}) remains intact in the index.
Note that the first prefactor in (\ref{MAEsol})
becomes trivial when we take the 4d limit.

By exploiting this fact, the results of $F_g\ (g\le 15)$
are packed into the following concise expression:
\[
Z^{E_8}
=\exp\left(\frac{E_{2*}}{24}\hbar^2\partial_\phi^2\right)
\left[
  \phi^{-5/2}
  \exp\sum_{k=1}^\infty
  c^{E_8}_k\left(\frac{\hbar}{\phi}\right)^{6 k}
\right],
\]
with
\[
c^{E_8}_1\Eqn{=}2525
 \left(\frac{E_{6*}}{2^{11}\cdot 3^4}\right),\nn\\
c^{E_8}_2\Eqn{=}1941020160
 \left(\frac{E_{6*}}{2^{11}\cdot 3^4}\right)^2,\nn\\
c^{E_8}_3\Eqn{=}26440099120581792
 \left(\frac{E_{6*}}{2^{11}\cdot 3^4}\right)^3,\nn\\
c^{E_8}_4\Eqn{=}2512057097259272539155456
 \left(\frac{E_{6*}}{2^{11}\cdot 3^4}\right)^4.
\]

Similarly, one can compute $Z^{H_0}$
by zooming in on the $H_0$ singularity at $z=\infty$.
This is done by first replacing the variables as
\[
\phi-\phi_0^{H_0}\to R\phi,\qquad
z\to R^{-1/5}z,\qquad
\hbar\to R\hbar
\]
and then taking the limit of $R\to 0$.
The mirror map (\ref{phiexpH0}) becomes
\[
\phi=\frac{6}{5z^5}.
\]
The BPS index for the $H_0$ SCFT is obtained as
\[
Z^{H_0}(\phi,\hbar)
=({\rm const.})
\lim_{R\to 0}Z^{E_8\oplus H_0}\left(R\phi+\phi_0^{H_0},R\hbar\right)
 \Bigm|_{z\sim \infty}.
\]
The results are given as follows:
\[
Z^{H_0}
=\exp\left(\frac{E_{2*}}{24}\hbar^2\partial_\phi^2\right)
\left[
  \phi^{-1/10}
  \exp\sum_{k=1}^\infty
  c^{H_0}_k\left(\frac{\hbar}{\phi}\right)^{6 k}
\right],
\]
with
\[
c^{H_0}_1\Eqn{=}2208217
 \left(\frac{E_{6*}}{2^{11}\cdot 3^4\cdot 5^7}\right),\nn\\
c^{H_0}_2\Eqn{=}85679149172703360
 \left(\frac{E_{6*}}{2^{11}\cdot 3^4\cdot 5^7}\right)^2,\nn\\
c^{H_0}_3\Eqn{=}76522745976613844587093143840
 \left(\frac{E_{6*}}{2^{11}\cdot 3^4\cdot 5^7}\right)^3,\nn\\
c^{H_0}_4\Eqn{=}516442282343494619890254623170666086528000
 \left(\frac{E_{6*}}{2^{11}\cdot 3^4\cdot 5^7}\right)^4.\quad
\]
%

\section{$E_7\oplus H_1$ case}

In this section we consider the compactification of the E-string
theory that realizes the $E_7$ and $H_1$ SCFTs and compute the BPS index.
Let us first recall the $E_7\oplus A_1$ reduction of
the global symmetry $E_8$ of the E-string theory.
This is achieved by setting the Wilson line parameters
as \cite{Eguchi:2002nx}
\[\label{mE7A1}
\bvec{m}=\left(0,0,0,0,0,0,\pi,\pi\right).
\]
In this case the general Seiberg--Witten curve
\cite{Eguchi:2002fc, Sakai:2011xg}
reduces to \cite{Sakai:2012ik}
\[\label{E7A1curve}
y^2=4x^3+\left(\varth_3^4+\varth_4^4\right)u^2x^2
+\left(\frac{\varth_3^4\varth_4^4}{4}u
  -\frac{16}{\varth_3^2\varth_4^2}\right)u^3x.
\]
Here $\varth_j\equiv\varth_j(0,\tau)$.
The $E_7$ singularity has already realized in this setting.
To obtain the $H_1$ singularity and make the complex structure
of the curve constant,
we set the complex structure of $T^2$ as
\[
\tau=\tau_\sharp:=\frac{1+i}{2}.
\]
This value is connected with $\tau=i$ by
an $\grp{SL}(2,\bbZ)$ transformation.\footnote{
We could study the present case
by adopting the value $\tau=i$.
This gives rise to different values of $E_{2n}(\tau)$.
In this case, however, the value of $\bvec{m}$
which gives the Seiberg--Witten curve
with $E_7$ and $H_1$ degenerations also changes.
(It changes to
the modular S-transform of (\ref{mE7A1}) \cite{Eguchi:2002nx}.)
Accordingly the form of the curve (\ref{E7A1curve})
and the normalization of $\phi$
are modified,
which compensates the changes of $E_{2n}(\tau)$
and eventually leads us to the same final results.
}
As in the last section, subscript $\sharp$ denotes
that the quantity is evaluated at $\tau=\tau_\sharp$.
We see that
\[
q_\sharp\Eqn{=}-e^{-\pi},\\[1ex]
E_{2\sharp}\Eqn{=}\frac{6}{\pi},\qquad
E_{4\sharp}=-12\frac{\Gamma(1/4)^8}{(2\pi)^6},\qquad
E_{6\sharp}=0,\\
\varth_{3\sharp}^4+\varth_{4\sharp}^4\Eqn{=}0,\qquad
\varth_{3\sharp}^2\varth_{4\sharp}^2
 =2\frac{\Gamma(1/4)^4}{(2\pi)^3}.
\]
The curve (\ref{E7A1curve}) then becomes
\[
y^2=4x^3
+\left(\frac{\Gamma(1/4)^8}{(2\pi)^6}u
  -8\frac{(2\pi)^3}{\Gamma(1/4)^4}
\right)u^3x.
\]

The mirror map is given as follows. The period is
\[
\omega
=\frac{1}{u\left(1-\frac{64}{\varth_{3\sharp}^6\varth_{4\sharp}^6u}
           \right)^{1/4}}
=\frac{\varth_{3\sharp}^6\varth_{4\sharp}^6}{64}\frac{z^4-1}{z^3},
\]
where this time we have defined $z$ as
\[
z:=
\left(1-\frac{64}{\varth_{3\sharp}^6\varth_{4\sharp}^6u}\right)^{-1/4}.
\]
The Higgs expectation value is given by
\[
\phi\Eqn{=}\int\omega du\nn\\
 \Eqn{=}4\int\frac{dz}{1-z^4}\\
 \Eqn{=}2\arctan z + 2\arctanh z+\phi_0^{E_7}.
\]
Near the $E_7$ singularity at $z=0$,
$\phi$ is expanded as
\[
\phi-\phi_0^{E_7}
 =4\sum_{k=0}^\infty \frac{z^{4k+1}}{4k+1}
 =4\left(z+\frac{z^5}{5}+\frac{z^{9}}{9}+\cdots\right).
\]
Near the $H_1$ singularity at $z=\infty$,
on the other hand, $\phi$ is expanded as
\[
\phi-\phi_0^{H_1}
 =4\sum_{k=0}^\infty\frac{1}{(4k+3)z^{4k+3}}
 =4\left(\frac{1}{3z^3}+\frac{1}{7z^{7}}
        +\frac{1}{11z^{11}}+\cdots\right).
\]

Using the explicit forms of $F_g\ (g\le 3)$ \cite{Sakai:2011xg},
one can immediately compute
$F^{E_7\oplus H_1}_g\ (g\le 3)$. The results are
\[
F^{E_7\oplus H_1}_0\Eqn{=}0,\\
F^{E_7\oplus H_1}_1\Eqn{=}\frac{1}{2}\ln\omega+\frac{1}{2}\phi
  -\frac{1}{2}
    \ln\left(-q_\sharp\prod_{k=1}^\infty(1-q_\sharp^k)^{12}\right)\nn\\
 \Eqn{=}
  \frac{1}{2}\ln\frac{z^4-1}{z^3}+\frac{1}{2}\phi
  +\frac{\pi}{4}-\frac{3}{2}\ln 2,\\[1ex]
F^{E_7\oplus H_1}_2
  \Eqn{=}
  \left(\frac{15-6z^4+7z^8}{1536z^{2}}
       -\frac{1}{96}\right)E_{2\sharp},\\[1ex]
F^{E_7\oplus H_1}_3
  \Eqn{=}\frac{(1-z^4)^2}{z^4}
  \left[\frac{15+22z^4+35z^8}{98304}E_{2\sharp}^2
  +\frac{9-11z^4+56z^8}{221184} E_{4\sharp}\right].
\]
By taking limits as in the last section, one obtains
the BPS index for the $E_7$ and $H_1$ SCFTs as
\[
Z^{E_7}
\Eqn{=}
 \exp\left(\frac{E_{2\sharp}}{24}\hbar^2\partial_\phi^2\right)
\left[
  \phi^{-3/2}
  \exp\sum_{k=1}^\infty
  c^{E_7}_k\left(\frac{\hbar}{\phi}\right)^{4 k}
\right],\\
Z^{H_1}
\Eqn{=}
 \exp\left(\frac{E_{2\sharp}}{24}\hbar^2\partial_\phi^2\right)
\left[
  \phi^{-1/6}
  \exp\sum_{k=1}^\infty
  c^{H_1}_k\left(\frac{\hbar}{\phi}\right)^{4 k}
\right]
\]
with
\[
c^{E_7}_1=\frac{E_{4\sharp}}{2^5\cdot 3},\qquad
c^{H_1}_1\Eqn{=}\frac{7E_{4\sharp}}{2^2\cdot 3^7}.
\]
To determine the invariants $c^{E_7}_k,c^{H_1}_k$ for $k\ge 2$,
we need more data of $F_g$ or $Z_n$.
We expect that $c_k^{E_7},\,c_k^{H_1}$ for general $k$ are given by
$E_{4\sharp}^k$ multiplied by some rational numbers,
as in the case of the $E_8$ and $H_0$ SCFTs.

\section{$E_6\oplus H_2$ case}

In this section we consider the compactification of the E-string
theory that realizes the $E_6$ and $H_2$ SCFTs and compute the BPS index.
Let us first recall the $E_6\oplus A_2$ reduction of
the global symmetry $E_8$ of the E-string theory.
This is achieved by setting the Wilson line parameters
as \cite{Eguchi:2002nx}
\[
\bvec{m}
 =\left(0,0,0,0,0,\frac{4\pi}{3},\frac{4\pi}{3},\frac{4\pi}{3}\right).
\]
The general Seiberg--Witten curve
\cite{Eguchi:2002fc, Sakai:2011xg}
reduces to \cite{Sakai:2012ik}
\[\label{E6A2curve}
y^2=
4x^3+3\alpha_{3}^2u^2x^2
+\frac{2}{3}\alpha_{3}\left(\beta_{3}u-\frac{27}{\beta_{3}}\right)u^3x
+\frac{1}{27}\left(\beta_{3}u-\frac{27}{\beta_{3}}\right)^2u^4,
\]
where
\[
\alpha_{3}\Eqn{:=}\sum_{(m,n)\in\bbZ^2}q^{m^2+n^2-mn}
            = \varth_3(0,2\tau)\varth_3(0,6\tau)
             +\varth_2(0,2\tau)\varth_2(0,6\tau),\nn\\
\beta_{3} \Eqn{:=} \frac{\eta(\tau)^9}{\eta(3\tau)^3}.
\]
The $E_6$ singularity has already realized in this setting.
To obtain the $H_2$ singularity and make the complex structure
of the curve constant, we set the complex structure of $T^2$ as
\[
\tau_\flat=\frac{1}{2}+\frac{i}{2\sqrt{3}}.
\]
This value is connected with $\tau=e^{2\pi i/3}$ by
an $\grp{SL}(2,\bbZ)$ transformation.
We let subscript $\flat$ denote
that the quantity is evaluated at $\tau=\tau_\flat$.
We see that
\[
q_\flat\Eqn{=}-e^{-\pi/\sqrt{3}},\\[1ex]
E_{2\flat}\Eqn{=}\frac{6\sqrt{3}}{\pi},\qquad
E_{4\flat}=0,\qquad
E_{6\flat}=-2^3\cdot 3^6\frac{\Gamma(1/3)^{18}}{(2\pi)^{12}},\\
\alpha_{3\flat}\Eqn{=}0,\qquad
\beta_{3\flat}=27\frac{\Gamma(1/3)^9}{(2\pi)^6}.
\]
The curve (\ref{E6A2curve}) then becomes
\[
y^2=
4x^3
+\frac{1}{27}\left(
 27\frac{\Gamma(1/3)^9}{(2\pi)^6}u
 -\frac{(2\pi)^6}{\Gamma(1/3)^9}\right)^2u^4.
\]

The mirror map is given as follows. The period is
\[
\omega
=\frac{1}{u\left(1-\frac{27}{\beta_{3\flat}^2 u}\right)^{1/3}}
=\frac{\beta_{3\flat}^2}{27}\frac{z^3-1}{z^2},
\]
where this time we have defined $z$ as
\[
z:=\left(1-\frac{27}{\beta_{3\flat}^2 u}\right)^{-1/3}.
\]
The Higgs expectation value is given by
\[
\phi\Eqn{=}\int\omega du\nn\\
 \Eqn{=}3\int\frac{dz}{1-z^3}\\
 \Eqn{=}-\ln(1-z)+\frac{1}{2}\ln(1+z+z^2)
 +\sqrt{3}\arctan\frac{1+2z}{\sqrt{3}}-\frac{\pi}{2\sqrt{3}}
 +\phi_0^{E_6}.
\]
Near the $E_6$ singularity at $z=0$, $\phi$ is expanded as
\[
\phi-\phi_0^{E_6}
 =3\sum_{k=0}^\infty \frac{z^{3k+1}}{3k+1}
 =3\left(z+\frac{z^4}{4}+\frac{z^{7}}{7}+\cdots\right).
\]
Near the $H_2$ singularity at $z=\infty$, on the other hand, 
$\phi$ is expanded as
\[
\phi-\phi_0^{H_2}
 =3\sum_{k=0}^\infty\frac{1}{(3k+2)z^{3k+2}}
 =3\left(\frac{1}{2z^2}+\frac{1}{5z^5}
        +\frac{1}{8z^8}+\cdots\right).
\]

Using the explicit forms of $F_g\ (g\le 3)$ \cite{Sakai:2011xg},
one can immediately compute
$F^{E_6\oplus H_2}_g\ (g\le 3)$. The results are
\[
F^{E_6\oplus H_2}_0\Eqn{=}0,\\
F^{E_6\oplus H_2}_1\Eqn{=}\frac{1}{2}\ln\omega+\frac{1}{2}\phi
  -\frac{1}{2}
    \ln\left(-q_\flat\prod_{k=1}^\infty(1-q_\flat^k)^{12}\right)\nn\\
 \Eqn{=}
  \frac{1}{2}\ln\frac{z^3-1}{z^2}+\frac{1}{2}\phi
  +\frac{\pi}{4\sqrt{3}}-\frac{3}{4}\ln 3,\\[1ex]
F^{E_6\oplus H_2}_2
  \Eqn{=}
  \left(\frac{8-4z^3+5z^6}{864z^{2}}
       -\frac{1}{96}\right)E_{2\flat},\\[1ex]
F^{E_6\oplus H_2}_3
  \Eqn{=}\frac{(1-z^3)^2(20+26z^3+35z^6)}{93312z^4}E_{2\flat}^2.
\]
The BPS index for the $E_6$ and $H_2$ SCFTs are
\[
Z^{E_6}
\Eqn{=}
 \exp\left(\frac{E_{2\flat}}{24}\hbar^2\partial_\phi^2\right)
\left[
  \phi^{-1}
  \exp\sum_{k=1}^\infty
  c^{E_6}_k\left(\frac{\hbar}{\phi}\right)^{6 k}
\right],\\
Z^{H_2}
\Eqn{=}
 \exp\left(\frac{E_{2\flat}}{24}\hbar^2\partial_\phi^2\right)
\left[
  \phi^{-1/4}
  \exp\sum_{k=1}^\infty
  c^{H_2}_k\left(\frac{\hbar}{\phi}\right)^{6 k}
\right].
\]
In contrast to the $E_8\oplus H_0$ case,
the holomorphic anomaly equation, the gap condition
and the data of elliptic genera
$Z_n\ (n\le 4)$ are not enough
to determine the invariants $c^{E_6}_k,\, c^{H_2}_k$.
To determine the first invariants $c^{E_6}_1,\, c^{H_2}_1$,
one needs slightly more data, e.g., the explicit form(s) of
$F^{E_6\oplus H_2}_4$ or $Z^{E_6\oplus H_2}_n$ with $n=5,6$.
We expect that $c_k^{E_6},\,c_k^{H_2}$ for general $k$ are given by
$E_{6\flat}^k$ multiplied by some rational numbers,
as in the case of the $E_8$ and $H_0$ SCFTs.

\vspace{3ex}

\begin{center}
  {\bf Acknowledgments}
\end{center}

This work was supported in part by
JSPS KAKENHI Grant Number 26400257,
JSPS Japan--Hungary and JSPS Japan--Russia
Research Cooperative Programs.

\vspace{3ex}


\renewcommand{\section}{\subsection}
\renewcommand{\refname}

\end{document}